\documentclass[a4paper,10pt,twoside]{cpc-hepnp}
\usepackage{CJK,upgreek,fancyhdr}
\usepackage{multicol}
\usepackage{graphicx}
\usepackage{booktabs}
\usepackage{amssymb,bm,mathrsfs,bbm,amscd}
\usepackage[tbtags]{amsmath}
\usepackage{lastpage}

\newcommand{\lambdabar}{{\mkern0.75mu\mathchar '26\mkern -9.75mu\lambda}}
\begin{document}
\begin{CJK*}{GB}{gbsn}

\fancyhead[c]{\small Chinese Physics C~~~Vol. xx, No. x (201x) xxxxxx}
\fancyfoot[C]{\small 010201-\thepage}

\footnotetext[0]{Received 31 April 2017}

\title{Thermonuclear $^{19}$F($p$,$\alpha_0$)$^{16}$O reaction rate\thanks{Supported by National Natural Science Foundation of China
(11490562, 11490560, 11675229) and by the National Key Research and Development Program of China (2016YFA0400503).}}

\author{%
      Jian-Jun He (何建军)$^{1;1)}$\email{hejianjun@nao.cas.cn}  \quad Ivano Lombardo$^{2)}$  \quad Daniele Dell'Aquila$^{3,4)}$ \\
      \quad Yi Xu (徐毅)$^{5,6)}$ \quad Li-Yong Zhang (张立勇)$^{1)}$ \quad Wei-Ping Liu (柳卫平)$^{7)}$
}
\maketitle

\address{%
$^1$Key Laboratory of Optical Astronomy, National Astronomical Observatories, Chinese Academy of Sciences, Beijing 100012, China\\
$^2$INFN - Sezione di Catania, via S. Sofia, I-95123, Catania, Italy\\
$^3$Dip. di Fisica, Univ. di Napoli Federico II, via Cintia, I-80126, Napoli, Italy\\
$^4$INFN - Sezione di Napoli, via Cintia, I-80126, Napoli, Italy\\
$^5$Extreme Light Infrastructure - Nuclear Physics, 30 Reactorului Street, P.O. Box MG-6, 077125 Magurele, jud. Ilfov, Romania\\
$^6$Nuclear Physics Institute, Czech Academy of Sciences, \v{R}e\v{z}, 25068, Czech Republic\\
$^7$China Institute of Atomic Energy, P. O. Box 275(10), Beijing 102413, China
}

\begin{abstract}
The thermonuclear $^{19}$F($p$,$\alpha_0$)$^{16}$O reaction rate in a temperature region of 0.007--10 GK has been derived by re-evaluating the
available experimental data, together with the low-energy theoretical $R$-matrix extrapolations. Our new rate deviates up to about 30\% compared
to the previous ones, although all rates are consistent within the uncertainties. At very low temperature (e.g. 0.01 GK) our reaction rate is
about 20\% smaller than the most recently published rate, because of a difference in the low energy extrapolated $S$-factor and a more accurate
estimate of the reduced mass entering in the calculation of the reaction rate. At temperatures above $\sim$1 GK, our rate is smaller, for instance,
by about 20\% around 1.75 GK, because we have re-evaluated in a meticulous way the previous data (Isoya et al., Nucl. Phys. 7, 116 (1958)). The
present interpretation is supported by the direct experimental data. The uncertainties of the present evaluated rate are estimated to be about 20\%
in the temperature region below 0.2 GK, which are mainly caused by the lack of low-energy experimental data and the large uncertainties of the
existing data. The asymptotic giant branch (AGB) star evolves at temperatures below 0.2 GK, where the $^{19}$F($p$,$\alpha$)$^{16}$O reaction
may play a very important role. However, the current accuracy of the reaction rate is insufficient to help to describe, in a careful way, for
the fluorine overabundances phenomenon observed in AGB stars. Precise cross section (or $S$ factor) data in the low energy region are therefore
mandatory for astrophysical nucleosynthesis studies.
\end{abstract}



\footnotetext[0]{\hspace*{-3mm}\raisebox{0.3ex}{$\scriptstyle\copyright$}2013
Chinese Physical Society and the Institute of High Energy Physics
of the Chinese Academy of Sciences and the Institute
of Modern Physics of the Chinese Academy of Sciences and IOP Publishing Ltd}%

\begin{multicols}{2}

\section{Introduction}
$^{19}$F is the unique stable fluorine isotope in nature. Its abundance is quite sensitive to the physical conditions of stars~\cite{luc11}.
The phenomenon of fluorine overabundances by factors of 800--8000 has been observed in R-Coronae-Borealis stars, providing evidence for the
fluorine synthesis in such hydrogen-deficient supergiants~\cite{pan08}.
In fact, $^{19}$F can be produced in the convective zone triggered by a thermal pulse in asymptotic giant branch (AGB) stars~\cite{cri09}, which
are the main contributors to the Galactic fluorine~\cite{jor92}. By so far, the astronomically observed fluorine overabundances could not be understood
by using current AGB models, and it seems that additional mixing effects should be involved~\cite{lug04}.
It shows that deep mixing phenomena in AGB stars could change the stellar outer-layer isotopic composition because of the proton capture reactions,
and affect the transported material~\cite{nol03,ser10,bus10}.
In this scenario, the main fluorine destruction reaction $^{19}$F($p$,$\alpha$)$^{16}$O possibly plays a role in modifying the fluorine surface
abundances~\cite{luc11,abi11}. As well, the hydrogen mixing is also important in the model of hydrogen-deficient post-AGB stars, and it can lead to estimates
of elemental abundances in better agreement with experimental findings~\cite{cla07}.

In nuclear physics aspects, thermonuclear $^{19}$F($p$,$\alpha$)$^{16}$O reaction rate is still not sufficiently accurate to address the fluorine overabundances
problem, especially the $^{19}$F($p$,$\alpha_0$)$^{16}$O rate in the low temperature region below 0.2 GK, where it dominates the total
$^{19}$F($p$,$\alpha$)$^{16}$O rate. Therefore, a detailed description of fluorine nucleosynthesis is still missing in despite of its crucial
importance.

Figure~\ref{fig1} shows scheme for the $^{19}$F($p$,$\alpha$)$^{16}$O reaction. It is well-known that this reaction takes place via three different types of
channels: ($p$,$\alpha_0$), ($p$,$\alpha_{\pi}$) and ($p$,$\alpha_{\gamma}$). Here after, the group of ($p$,$\alpha_2$), ($p$,$\alpha_3$) and
($p$,$\alpha_4$) accompanying with the $\gamma$ transitions of $\gamma_2$, $\gamma_3$ and $\gamma_4$, is referred to as the ($p$,$\alpha_{\gamma}$) channel.
In this work, we have re-evaluated the cross section data of $^{19}$F($p$,$\alpha_0$)$^{16}$O reaction in the center-of-mass ($c.m.$) energy region
up to 10 MeV. These data are sufficient to account for thermonuclear $^{19}$F($p$,$\alpha_0$)$^{16}$O reaction rate up to a temperature of 10 GK.
Together with the low-energy theoretical predictions for the $S$ factors, a new reaction rate has been derived in a temperature region of 0.007--10 GK.
Results concerning the other two reaction channels will be the subject of forthcoming papers.

\begin{center}
\includegraphics[width=7.5cm]{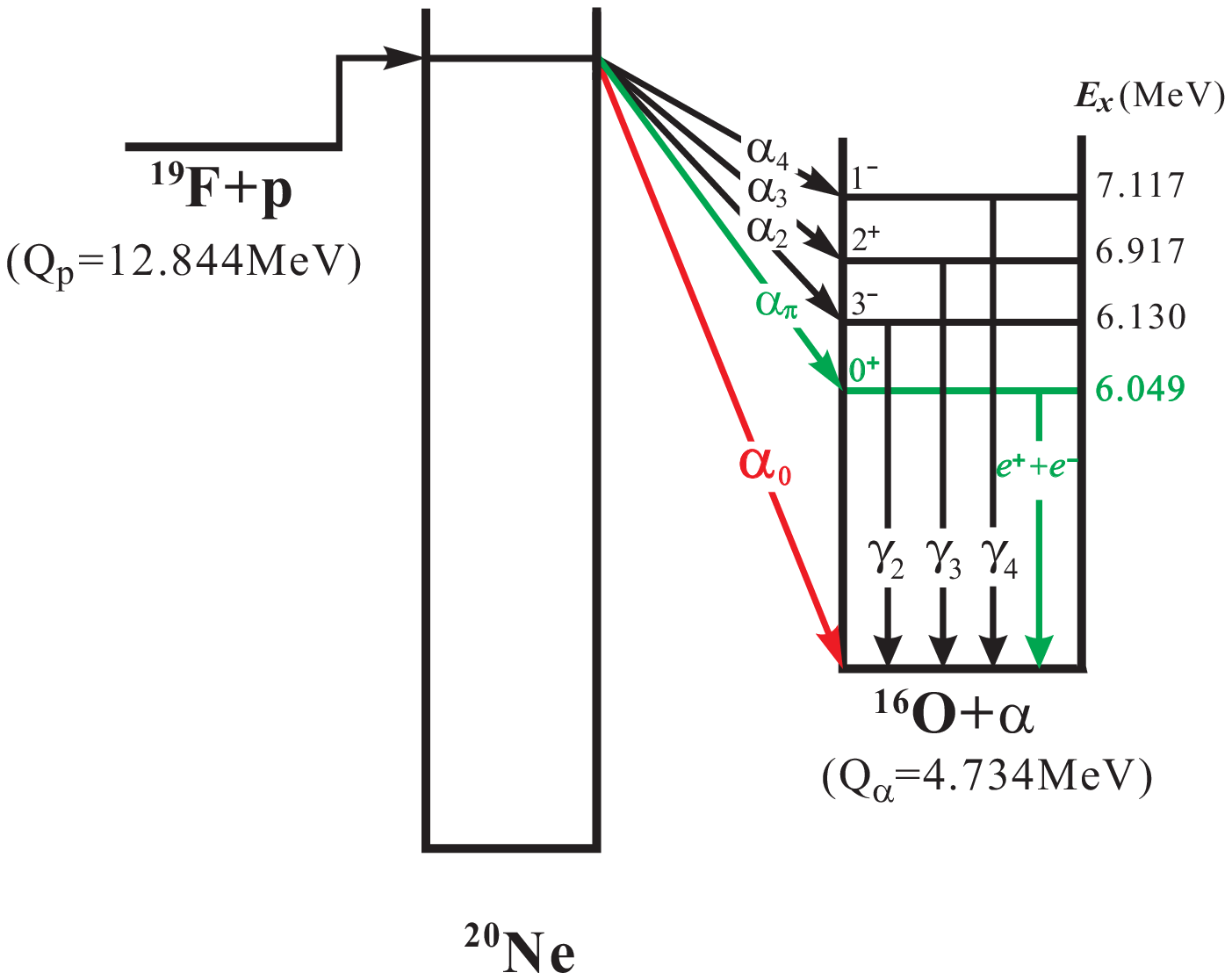}
\figcaption{Scheme of the $^{19}$F($p$,$\alpha$)$^{16}$O reaction.}
\label{fig1}
\end{center}

\section{NACRE compilation}
In the \emph{Nuclear Astrophysics Compilation of Reaction Rates} (NACRE)\footnote{http://pntpm.ulb.ac.be/Nacre/nacre.htm}~\cite{ang99}, the
$^{19}$F($p$,$\alpha_0$)$^{16}$O astrophysical $S(E)$-factors within $E_{c.m.}$=0.1--10 MeV were recommended on the basis of several
works~\cite{cla57,bre59,car74,cuz80,iso58,mor66,war63}, where the lowest direct energy point is close to $E_{c.m.}$=461 keV~\cite{bre59}. Figure~\ref{fig2}
shows the NACRE compiled $S$-factor data in a linear scale, where the discrepancies between different data sets can be clearly appreciated.
Three major discrepancies need to be pointed out: 1) in the $E_{c.m.}$=1.6--2.5 MeV region, CLA57~\cite{cla57} data are different from those of
CUZ80~\cite{cuz80}; 2) the resonance energy of the $E_{c.m.}$=1.3 MeV maximum in the cross section is reported to be located at 1.289 MeV in ISO58~\cite{iso58}
and 1.302 MeV in CLA57, with about 13 keV deviation; 3) BRE59~\cite{bre59} data are systematically larger than those of ISO58.

\begin{center}
\includegraphics[width=8.6cm]{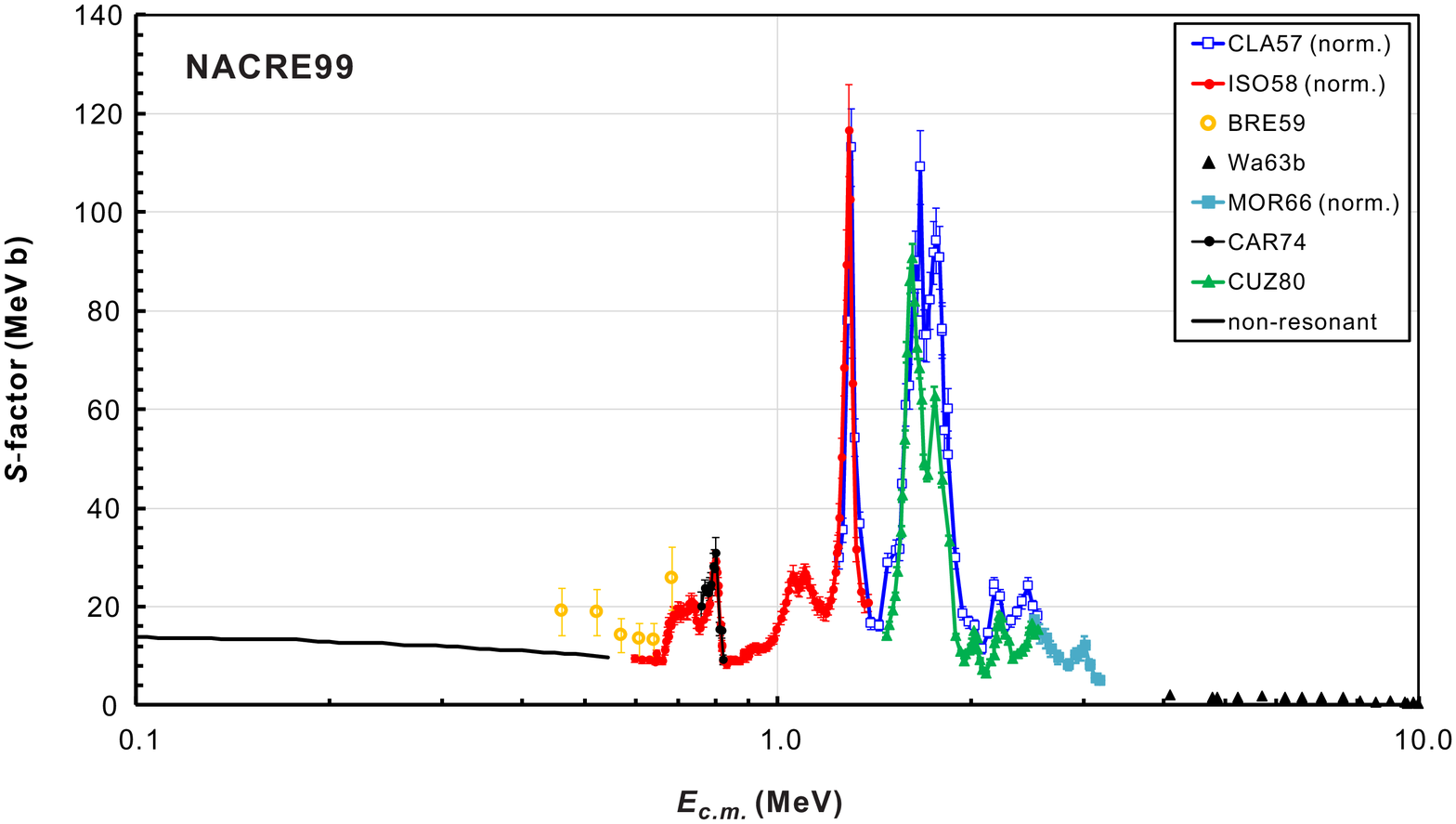}
\vspace{-1mm}
\figcaption{Astrophysical $S$ factors of the $^{19}$F($p$,$\alpha_0$)$^{16}$O reaction adopted in the NACRE compilation~\cite{ang99}.}
\label{fig2}
\end{center}

\section{Data after NACRE}
Lombardo et al. reported new direct measurement data~\cite{lom13,lom15} on the $^{19}$F($p$,$\alpha_0$)$^{16}$O reaction in the energy region of
$E_{c.m.}$=0.18--1 MeV. Figure~\ref{fig3} shows the NACRE data together with the new measurements for the $^{19}$F($p$,$\alpha_0$)$^{16}$O, where
data in the energy region above 1 MeV are not shown for clarity reasons. Here, the extrapolated low-energy non-resonant curves shown in
Figs.~\ref{fig2} and \ref{fig3} are taken from NACRE.
It should be noted that:
(1) LOM13~\cite{lom13} data are systematically larger than those of ISO58 below $\sim$0.75 MeV, but smaller above $\sim$0.85 MeV;
(2) LOM13 and LOM15~\cite{lom15} data are consistent with BRE59 data within uncertainties, but the latter has very large uncertainties.

La Cognata et al. reported indirect Trojan horse method (THM) results of COG11~\cite{cog11} and COG15~\cite{cog15} on this reaction; staring from the
experimentally determined resonance properties, the $S$ factor was deduced by $R$-matrix calculations. At temperatures around 0.1 GK, their rate is about
70\% larger than the NACRE one, and beyond the previous uncertainties~\cite{ang99}. Such difference was owing to the 113 keV resonance. But, their energy
resolution achieved was still not enough for a good separation between adjacent resonances. Just recently, a high-resolution THM experiment of
IND17~\cite{ind17} was performed and observed the 251 keV broad resonance clearly; by involving this broad resonance, they obtained a relatively higher
$S$-factor than that of COG15. However, the indirectly measured $S$ factors of IND17 are still lower than the directly measured data of LOM15, although
they are in agreement within the relatively large uncertainties (as shown in the following Fig.~\ref{fig8}).

\begin{center}
\includegraphics[width=8.6cm]{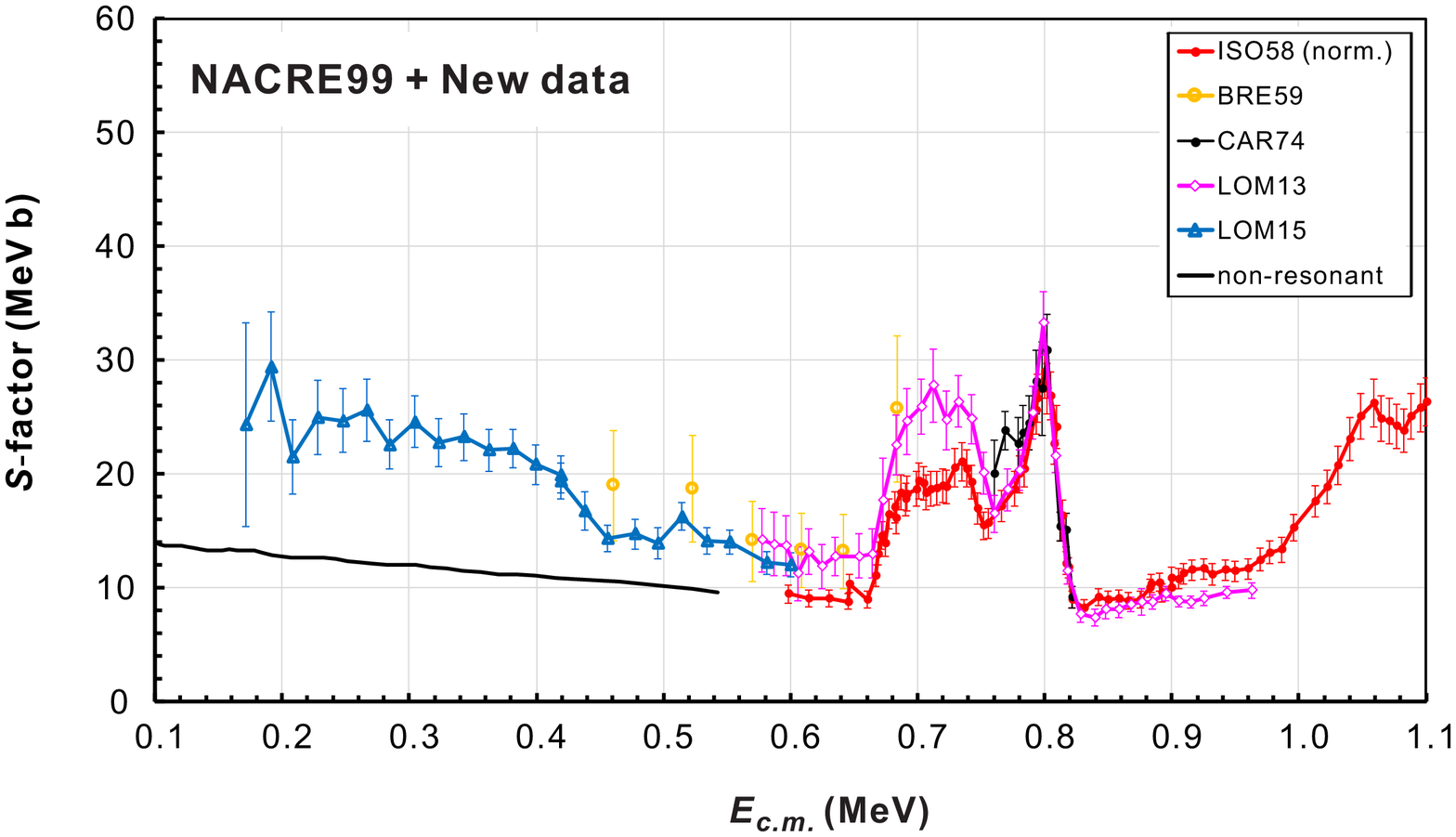}
\vspace{-1mm}
\figcaption{Part of astrophysical $S$ factors of the $^{19}$F($p$,$\alpha_0$)$^{16}$O reaction. It includes the data evaluated in the NACRE compilation~\cite{ang99}
and new direct measurement data~\cite{lom13,lom15}. Here, for clarity, the data within $E_{c.m.}$=1--10 MeV
region are not shown repeatedly, which are exactly the same as in Fig.~\ref{fig2}.}  
\label{fig3}
\end{center}

\section{Present evaluation}
In this work, we have extracted the experimental data or theoretical curves from the figures in the literature by using the \emph{GetData Graph
Digitizer} program\footnote{http://getdata-graph-digitizer.com/} (hereafter referred to as ``GetData"). Some data are also taken from the
\emph{Experimental Nuclear Reaction Data} (EXFOR) library\footnote{http://www.nndc.bnl.gov/exfor/exfor.htm}. We firstly digitized or deduced the
$^{19}$F($p$,$\alpha_0$)$^{16}$O cross section data, and then converted to the astrophysical $S$ factors by~\cite{rol88},
\begin{eqnarray}
\sigma(E)=\frac{1}{E}\mathrm{exp}(-2\pi\eta)S(E).
\label{eq1}
\end{eqnarray}
The quantity $\eta$ is called the Sommerfeld parameter and defined as $\eta$=$\frac{Z_1Z_2e^2}{\hbar v}$. In numerical units, the exponent is
2$\pi\eta$=31.29$Z_1Z_2$$\sqrt{\mu/E}$, where the center-of-mass energy $E$ is given in units of keV and the reduced mass $\mu$ is in amu. Here,
quantity $\mathrm{exp}(-2\pi\eta)$ is the Coulomb barrier penetration probability.

\subsection{Astrophysical $S$ factors}
The astrophysical $S$ factors have been evaluated in the $E_{c.m.}$=0.1--3.2 MeV region based on the up-to-date experimental data shown in Fig.~\ref{fig4}.
The higher energy `WA63b' data~\cite{war63} shown in Fig.~\ref{fig2} are adopted in the present evaluation. The low energy region of data is expanded
in Fig.~\ref{fig5} for clarity. We will discuss the details of our re-evaluation procedure of available data in the following subsections. It should be
noted that the solid lines connecting the data points shown in the following Figs.~2--10 are intended only as a guide for the eye. Here, the
uncertainties of BRE59, MOR66~\cite{mor66} and CAR74~\cite{car74} data are taken from NACRE~\cite{ang99}; those of LOM13 and LOM15 data are taken from Refs.~\cite{lom13,lom15}
including statistical plus systematical errors; NACRE assumed 3\% for CUZ80, while we digitize the errors from figure 3 of CUZ80; NACRE assumed 7\% for
ISO58 data, and we assume 10\% for these data relative to LOM13 data; NACRE assumed 7\% for CLA57 data, and we assume about 12\% for these data relative to ISO58 data.

\begin{center}
\includegraphics[width=8.6cm]{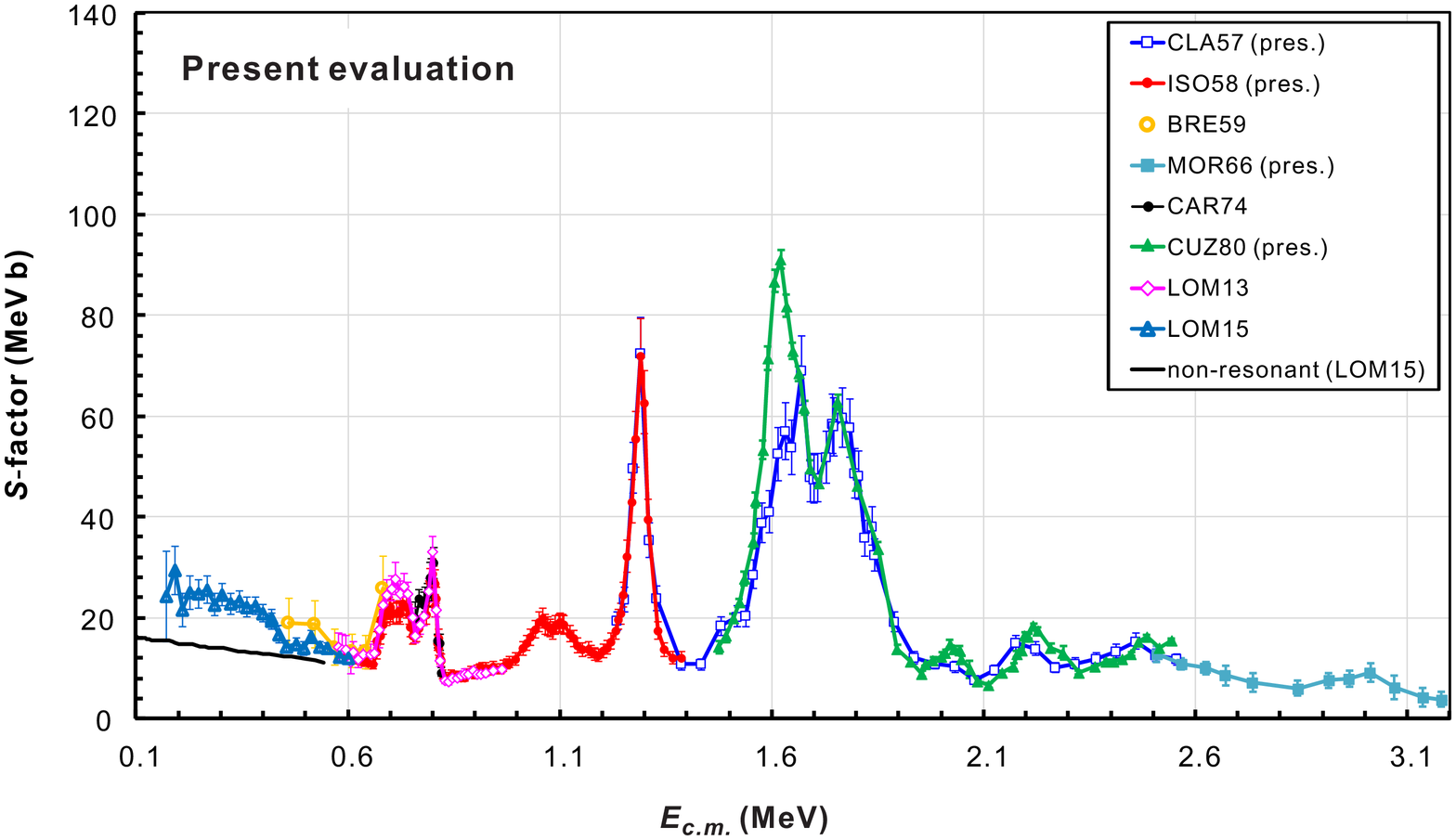}
\vspace{-1mm}
\figcaption{Present evaluated astrophysical $S$ factors of the $^{19}$F($p$,$\alpha_0$)$^{16}$O reaction. Data are taken from CLA57~\cite{cla57},
ISO58~\cite{iso58}, BRE59~\cite{bre59}, MOR66~\cite{mor66}, CAR74~\cite{car74}, CUZ80~\cite{cuz80}, LOM13~\cite{lom13} and LOM15~\cite{lom15}. The
theoretical non-resonant curve is taken from COG15~\cite{cog15} (i.e., that of NACRE by a scaling factor of 1.16). We have re-evaluated the CLA57, ISO58,
MOR66 and CUZ80 data which are indicated by `(pres.)' in the corresponding legends. It should be noted that the BRE59 data are not used in the present
reaction rate calculations due to their large uncertainties. Please see text for details.}
\label{fig4}
\end{center}

\begin{center}
\includegraphics[width=8.6cm]{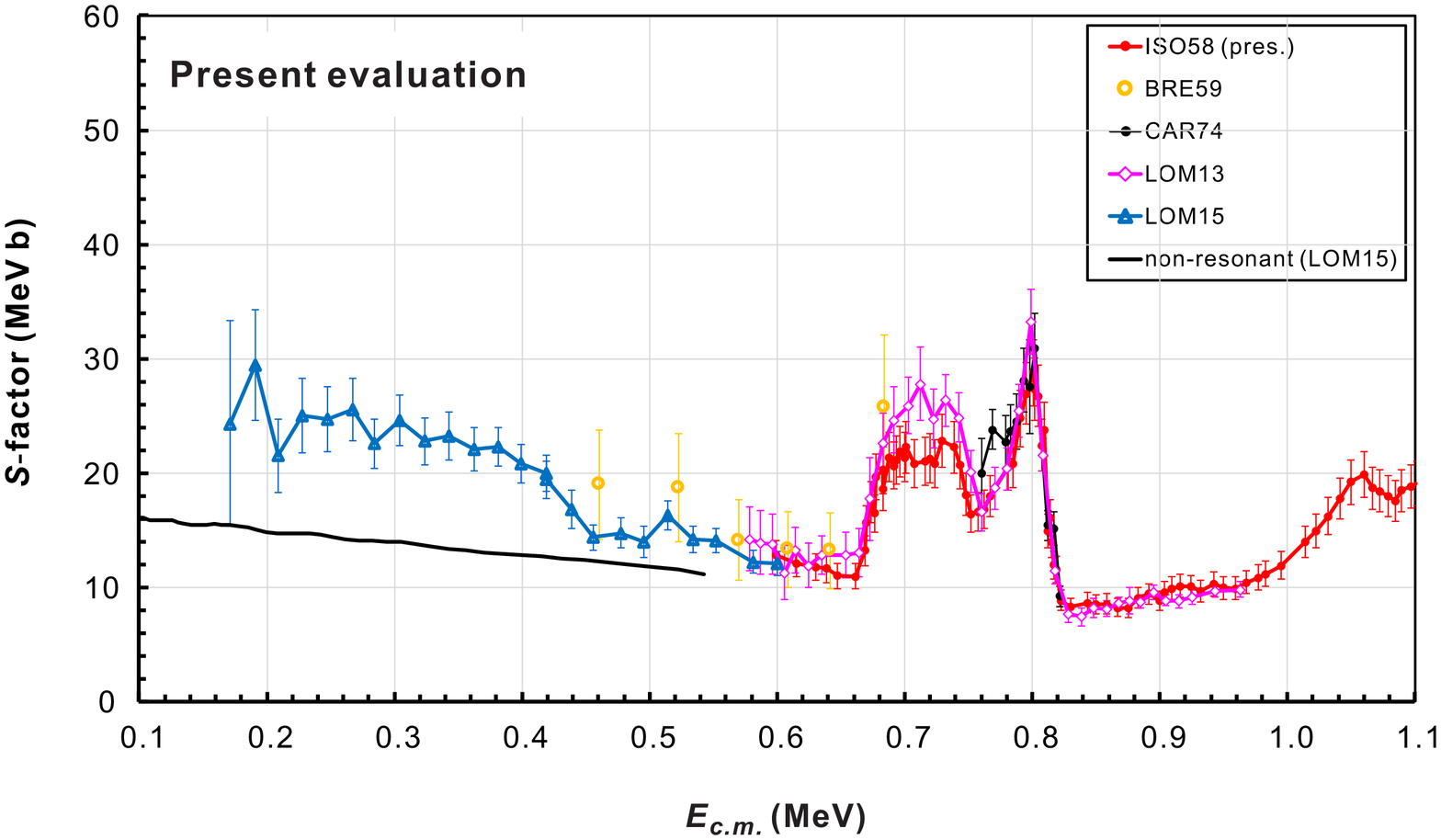}
\vspace{-1mm}
\figcaption{Similar caption as in Fig.~\ref{fig4}.}
\label{fig5}
\end{center}

\subsubsection{ISO58 data}
\noindent
The ISO58~\cite{iso58} data evaluated in NACRE are systematically smaller than the BRE59 and LOM13 data below $\sim$0.75 MeV (see Fig.~\ref{fig3}),
as already mentioned above. In order to find a possible explanation of such discrepancy, we have checked the $S$ factors of ISO58 taken from the NACRE
website and the Legendre polynomial coefficients in Ref.~\cite{iso58} carefully. Usually, the differential cross section can be reproduced by a
Legendre polynomial expansion:
\begin{eqnarray}
\frac{d\sigma}{d\Omega}(\theta)=\sum_{n} B_n P_n (\mathrm{cos}\theta).
\label{eq2}
\end{eqnarray}
In this frame, the total cross section can be calculated as $\sigma_\mathrm{tot}=4\pi B_0$. However, ISO58 expressed their angular distribution by a
different equation:
\begin{eqnarray}
\frac{d\sigma}{d\Omega}(\theta)=\frac{\lambdabar^2}{8}\sum_{n} b_n P_n (\mathrm{cos}\theta),
\label{eq3}
\end{eqnarray}
where the additional parameter $\lambdabar^2$ is inversely proportional to the $E_{c.m.}$ energy.
In NACRE, the relative cross sections of ISO58 were normalized to $\sigma$=42 mb at the 1.3 MeV resonance. By multiplying the $b_0$ data (taken from Fig. 4 in
Ref.~\cite{iso58}) by a factor of 2.97$\times$10$^{-4}$, the cross section at the 1.290 MeV resonance peaks at 42 mb, and also 2.97$\times$10$^{-4}$$\times$$b_0$
reproduces almost perfectly the ISO58 data evaluated by NACRE in the whole energy range. Therefore, we speculated that NACRE evaluated the ISO58 data by
the relation of 2.97$\times$10$^{-4}$$\times$$b_0$. In fact, the integrated cross section cannot be estimated by a simple scaling of the $b_0$ data,
and we have to take explicitly into account the energy dependence of $\lambdabar^2$ reported in Eq.~\ref{eq3}. We performed such a procedure and obtained
a new estimate of the integrated cross section starting from the $b_0$ data of ISO58. In Fig.~\ref{fig6}
we show the comparison between our new evaluation of ISO58 data (ISO58 (Corrected), in black) and the previous NACRE evaluation (ISO58 (NACRE), in
light blue). Significant differences appear at the two edges, i.e., the energy regions far away from 1 MeV, and it implies that the energy dependence
correction in $\lambdabar^2$ has considerable impact on the evaluated cross sections.
Finally, the presently evaluated ``ISO58 (pres.)" data, which are obtained by multiplying the ``ISO58 (Corrected)" data with a normalization factor of 0.8,
are consistent with those LOM13, LOM15 and BRE59 data in the whole energy range, as seen in Figs.~\ref{fig4} and \ref{fig5}. It shows that the
procedure here adopted to extract the cross section starting from the ISO58 $b_0$ data removes the discrepancies between various data set previously noticed
in the $E_{c.m.}$=0.6--1 MeV range.

In the present work, the peak cross section of the 1.3 MeV resonance is evaluated to be (26.0$\pm$2.6) mb based on the ISO58 data, a value quite lower
than the 42 mb value adopted by NACRE. In fact, there are no absolute cross section values reported in the published literature for this resonant peak.
Only Ref.~\cite{cuz80a} reported a value of 29 mb (with about 15\% total uncertainty), which agrees very well with the present value.

\begin{center}
\includegraphics[width=8cm]{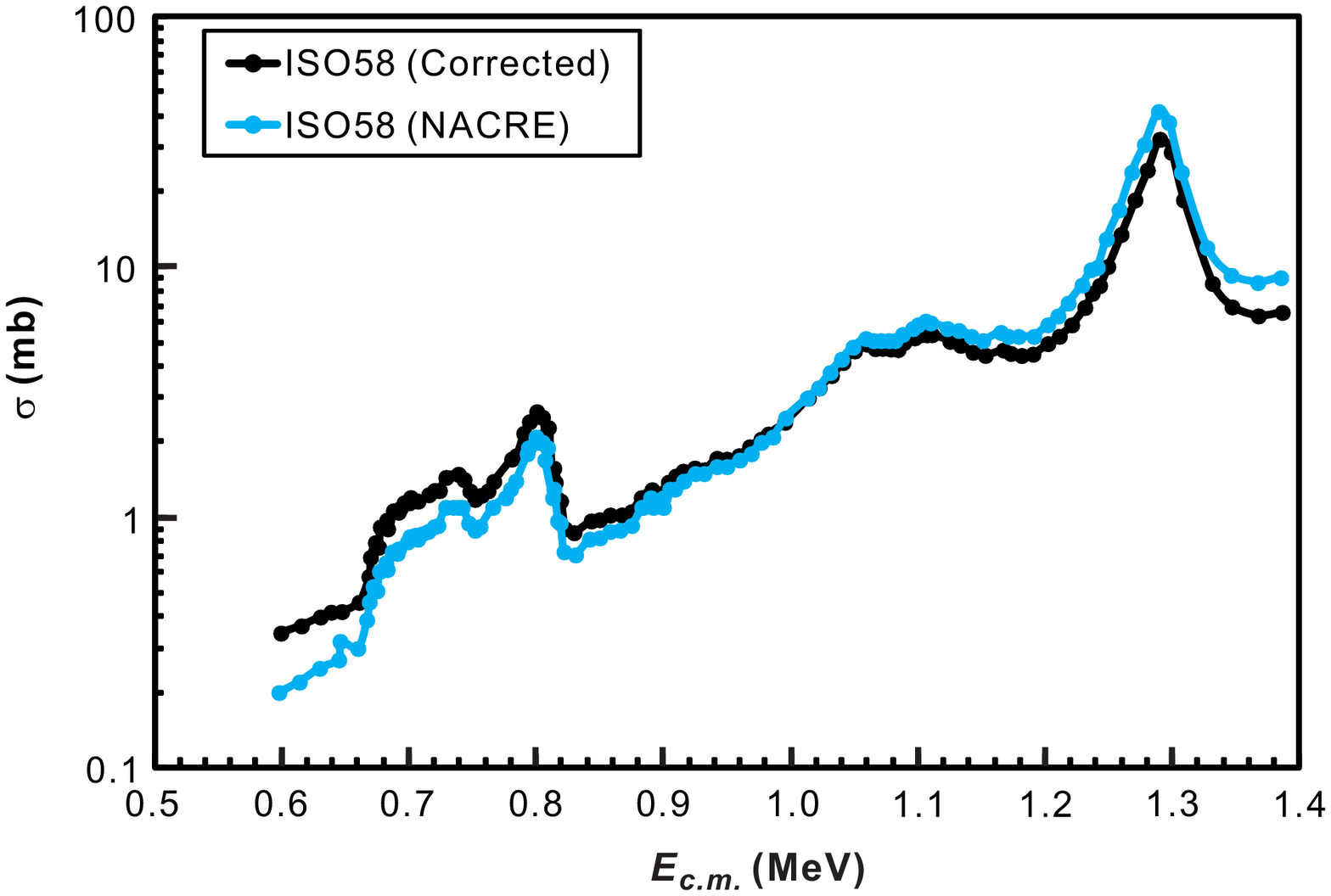}
\vspace{2mm}
\figcaption{Cross sections of the $^{19}$F($p$,$\alpha_0$)$^{16}$O reaction calculated based on the $b_0$ data of ISO58~\cite{iso58}. Here, the
``ISO58 (Corrected)" data multiplied by a normalization factor of 0.8 equal to the presently evaluated ``ISO58 (pres.)" data as shown in
Figs.~\ref{fig4} and \ref{fig5}.}
\label{fig6}
\end{center}

\subsubsection{CLA57 data}
In CLA57~\cite{cla57}, the yield of the ground state alpha particles from the $^{19}$F($p$,$\alpha_0$)$^{16}$O reaction was studied in a proton energy
range going from 1.3 to 2.7 MeV. The authors analyzed the observed angular distributions in terms of Legendre polynomial expansion (Eq.~\ref{eq1}),
and reported the trend of the coefficients as a function of energy. We have obtained the NACRE $S$-factor data from the NACRE website, and the Legendre
polynomial coefficient $a_0$ of Fig. 4 in Ref.~\cite{cla57} by GetData. We show the data corresponding to 2.51$\times$10$^{-2}$$\times$$a_0$ (labelled
as ``GetData") in Fig.~\ref{fig7} as red dots. We find that they are consistent with the NACRE evaluated ones where the relative cross sections of CLA57
were normalized to $\sigma$=42 mb at the 1.3 MeV resonance. However, the two energy scales are slightly different (especially at lower energies, where
their difference amounts to about 10 keV in the 1.3 MeV region). The present energy (``GetData") scale can match that of the ISO58 data better. In order
to match the present CLA57 data (``GetData") with the ISO58 data evaluated above, the former was multiplied by a factor of 0.63 in our final evaluation
(labelled as ``CLA57 (pres.)" in Fig.~\ref{fig4}). In Fig.~\ref{fig4}), the shapes of ``ISO58 (pres.)" and ``CLA57 (pres.)" are matched very well around
the 1.3 MeV resonance, where the values of the evaluated $S$ factor are 72.1 MeV$\cdot$b for ISO58 at 1.290 MeV, and 72.4 MeV$\cdot$b for CLA57 at 1.292 MeV,
respectively. The peak cross sections for six resonances listed in Table IV in Ref.~\cite{cla57} are also shown in Fig.~\ref{fig7} for comparison.
These data were determined relative to the known $^{19}$F($p$,$\alpha_\gamma$)$^{16}$O cross section. It shows they are roughly consistent with the
NACRE and present results within their large uncertainties, except two data points at 2.01 and 2.45 MeV. But, most of these data are much larger than
the present evaluation if considering the above factor of 0.63 for ``GetData" in Fig.~\ref{fig7}. Therefore, we conclude that these peak cross sections
data listed in Table IV in Ref.~\cite{cla57} are unreliable.

In fact, CLA57 derived a value of 46 mb at the $E_p$=1.358 MeV resonance by normalizing their results to a previously uncertain ($p$,$\alpha_\gamma$)
value at the 1.372 MeV resonance (i.e., 300 mb estimated by Streib et al.~\cite{str41}). By considering the above normalization factor of 0.63,
a value of 29 mb (=46$\times$0.63) is obtained, which is consistent with the present evaluated value of 26 mb.

\begin{center}
\includegraphics[width=8cm]{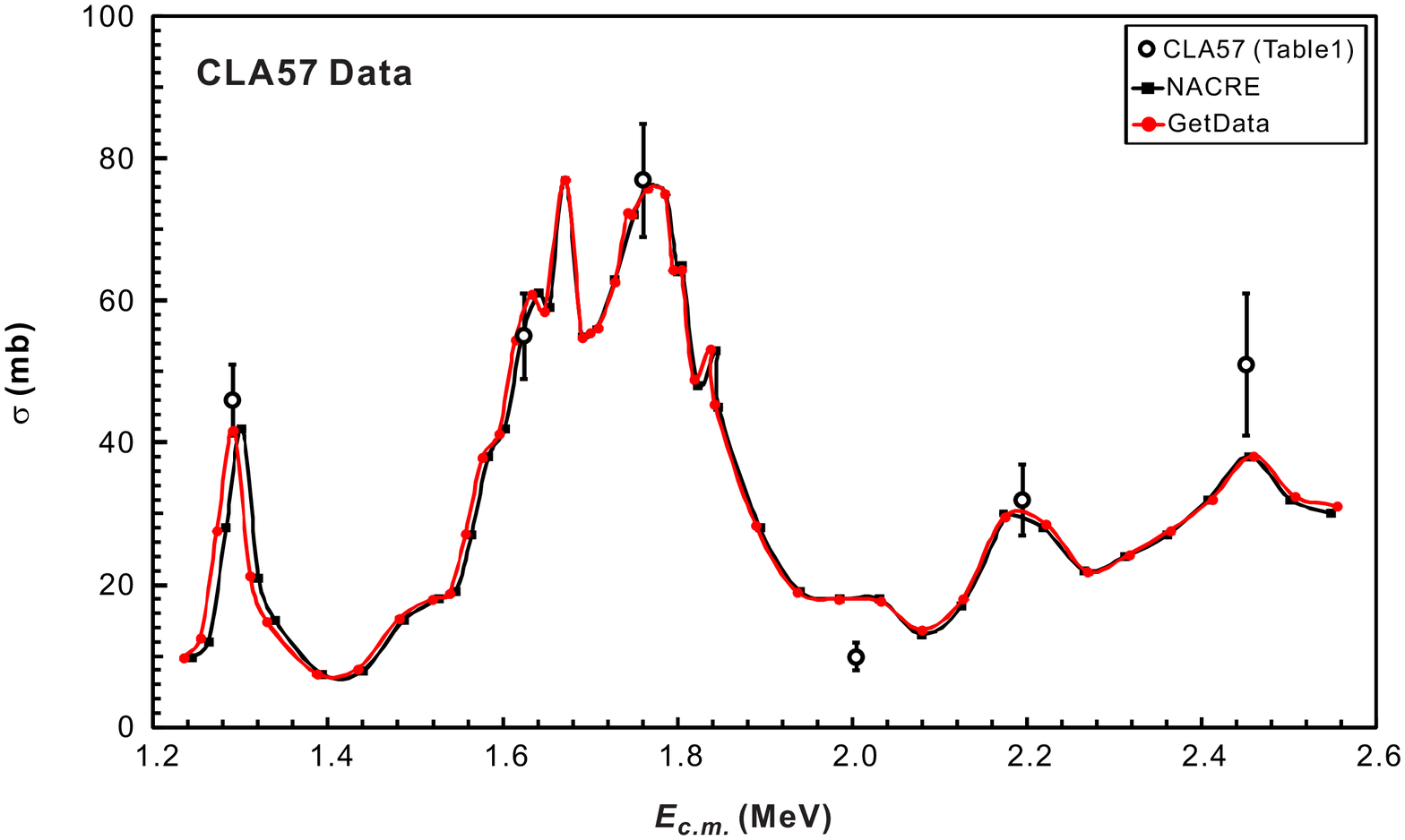}
\vspace{2mm}
\figcaption{Cross sections of the $^{19}$F($p$,$\alpha_0$)$^{16}$O reaction starting from the data reported by CLA57~\cite{cla57}.}
\label{fig7}
\end{center}

\subsubsection{CUZ80 data}
In the NACRE compilation, the data of Fig. 3 in CUZ80~\cite{cuz80} were digitized as shown in Figs.~\ref{fig2} and \ref{fig3}. However, they only
simply adopted about 3\% uncertainty on the data. In this work, we adopted the NACRE evaluated data, while the associated uncertainties were digitized
from the Fig. 3 in CUZ80. Uncertainties vary, depending on the energy, from about 2\% up to 20\%. The present evaluation is indicated as ``CUZ80 (pres.)"
in Fig.~\ref{fig4}. In the $E_{c.m.}$=1.52--1.65 MeV region, the ``CUZ80 (pres.)" data are considerably different form those of ``CLA57 (pres.)".
Therefore, new experiments are needed to clarify this discrepancy.

\subsubsection{MOR66 data}
In MOR66~\cite{mor66}, the coefficients of the Legendre polynomials were obtained at six energy points as listed in their Table I. By using the
coefficient $A_0$, NACRE normalized the data of MOR66 at 2.507 MeV to $\sigma$=28 mb, the averaged value of CLA57 and CUZ80 (see Fig.~\ref{fig2}).
In order to match the ``CLA57 (pres.)" data, we have normalized the MOR66 data at 2.507 MeV to $\sigma$=20.8 mb labelled as ``MOR66 (pres.)" in Fig.~\ref{fig4}.

\subsubsection{Low-energy extrapolation}
In NACRE, a non-resonant contribution was calculated below 0.46 MeV for $s$-wave capture with the procedure described in Ref.~\cite{cha50}, and then
adjusted to the lower experimental points in the 0.46$\leq$$E_{c.m.}$$\leq$0.60 MeV range. This non-resonant contribution matches well the
old NACRE ``ISO58 (norm.)" data as shown in Fig.~\ref{fig2}. In this work, we have adopted the non-resonant contribution fitted in the $R$-matrix
calculations of LOM15, i.e., the NACRE non-resonant contribution with a scaling factor of 1.16, as shown in Figs.~\ref{fig4} and \ref{fig5}.

In addition, the low-energy unpublished experimental data and theoretical predictions for the $^{19}$F($p$,$\alpha_0$)$^{16}$O reaction have been reviewed
in Ref.~\cite{wie99}. In the unpublished thesis work of LOR78~\cite{lor78}, differential cross sections were measured in the energy range between
$E_p$=0.14--0.90 MeV at the two angles $\theta_\mathrm{lab}$=90$^\circ$, 135$^\circ$, respectively. Relative angular distributions were measured at four proton
energies: 250, 350, 450 and 550 keV, respectively. The astrophysical $S$ factor was parameterized in the analytical form~\cite{rol88}
\begin{eqnarray}
S(E)=S(0)+S'(0)E+\frac{1}{2}S''(0)E^2,
\label{eq4}
\end{eqnarray}
with $S$(0)=3.77 MeV$\cdot$b, $S'$(0)=-5.13 b and $S''$(0)=90.75 b$\cdot$MeV$^{-1}$, by simply assuming $\sigma_\mathrm{tot}=4\pi \frac{d\sigma}{d\Omega}$(90$^\circ$).
Later on, HER91~\cite{her91} and YAM93~\cite{yam93} independently performed zero and finite-range Distorted Wave Born Approximation (DWBA) analysis of
the LOR78 data. Two calculated astrophysical $S$ factors agree within $\sim$15\%. Based on the predicted angular distribution, HER91 quoted a $S$ factor
($S$(0)=8.755 MeV$\cdot$b, $S'$(0)=-3.48 b and $S''$(0)=20.1 MeV$^{-1}$b) about a factor of two larger than the LOR78 one at low energies as shown in Fig.~\ref{fig8}.
As commented in NACRE, HER91 and YAM93 were focused mainly in the relative energy dependence of the cross section without accurate check on
the absolute cross sections which may be underestimated by a factor of 2. In fact, the underestimation of LOR78 data can be obviously seen in the following
Fig.~\ref{fig10}. Therefore, it seems reasonable that the unpublished LOR78 data were not included in the NACRE compilation.

Figure~\ref{fig8} shows the comparison between different predictions. It can be seen that HER91 result is still about a factor of 2 smaller than the presently
re-evaluated non-resonant contribution. In addition, the $R$-matrix results of LOM15 based on direct experimental data, as well as those of COG15 and IND17 based
on indirect THM data are also shown in Fig.~\ref{fig8}. It shows that the recent result of IND17 is quite close to that of LOM15, except in the energy region
around 0.2$\sim$0.4 MeV, although both results are roughly consistent within the large uncertainties. For clarity, only centroid value of IND17 is shown, and
actually an uncertainty of 16\% was assumed in their work. As evident from the figure, we are still lack of the experimental data in the energy region below
0.2 MeV, and the accuracy of the existing data around 0.2 MeV is not sufficient yet. Therefore, precise direct cross-section measurements are of great importance
to describe proton-induced fluorine destruction in astrophysical nucleosynthesis studies.

\begin{center}
\includegraphics[width=8.6cm]{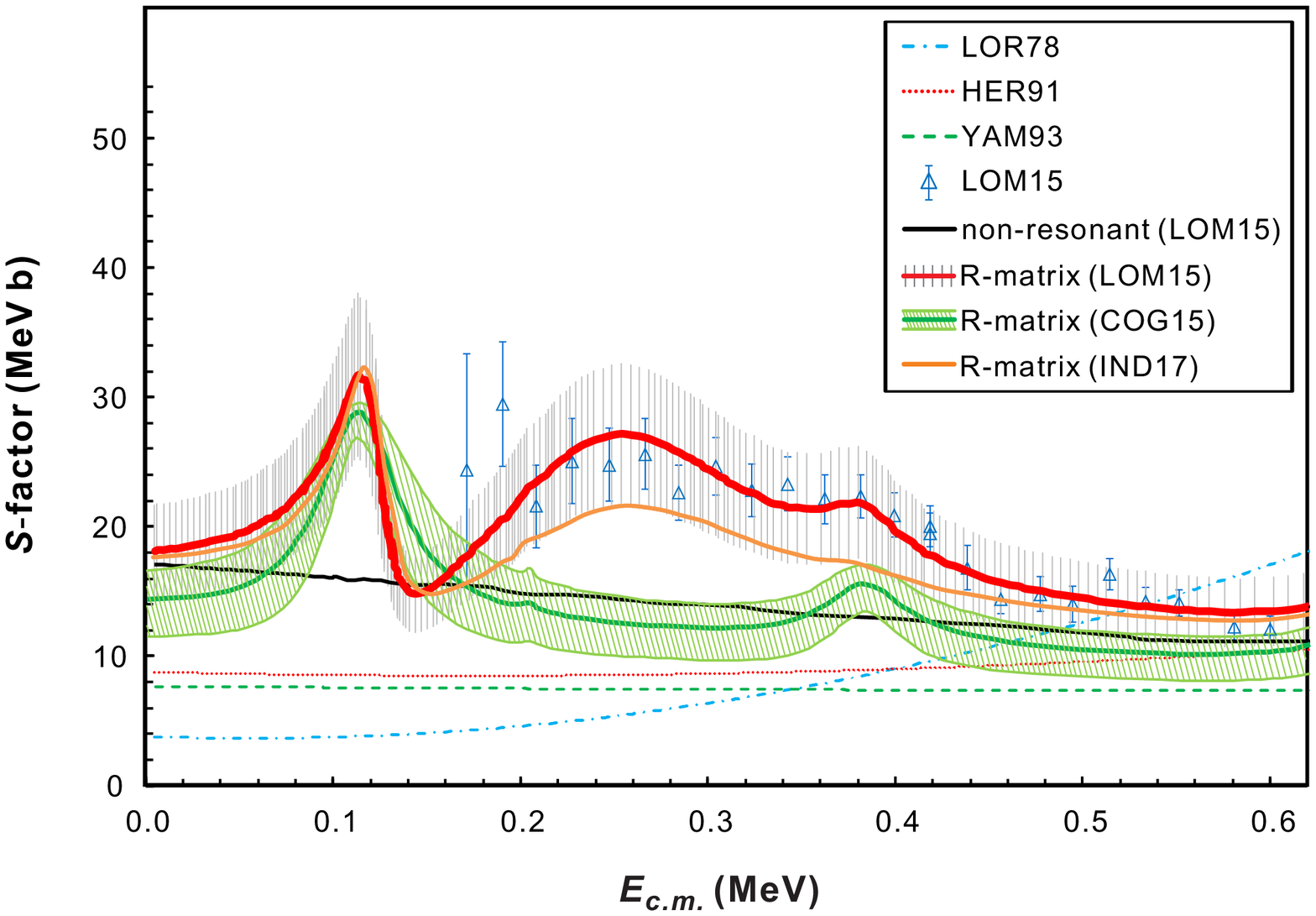}
\vspace{-0mm}
\figcaption{Low-energy astrophysical $S$ factors of the $^{19}$F($p$,$\alpha_0$)$^{16}$O reaction. The non-resonant predictions (LOR78~\cite{lor78},
HER91~\cite{her91}, YAM93~\cite{yam93} and LOM15~\cite{lom15}) and $R$-matrix results (COG15~\cite{cog15}, IND17~\cite{ind17}, and LOM15) are
shown for comparison.}
\label{fig8}
\end{center}

\subsection{Angular distribution}
In general, experimentally observed angular distributions can be fitted in two alternative ways: (1) the Legendre polynomials by Eq.~\ref{eq2} expressed
above, (2) the cosine polynomials, which can be expressed as
\begin{eqnarray}
\frac{d\sigma}{d\Omega}(\theta)=\sum_{n} A_n \mathrm{cos}^n\theta.
\label{eq5}
\end{eqnarray}
It can be easily shown that the total cross section can be deduced by the differential cross section at $\theta$=90$^\circ$ and the presently defined
angular distribution factor $f$ with the equation:
\begin{eqnarray}
\sigma_\mathrm{tot}=4\pi \times \frac{d\sigma}{d\Omega}(90^\circ) \times f.
\label{eq6}
\end{eqnarray}
The factor $f$ can be calculated with the coefficients of Legendre polynomials $B_i$ (up to 4$^\mathrm{th}$ order) by
\begin{eqnarray}
f = 1/\left(1 - \frac{B_2}{2B_0}+\frac{3B_4}{8B_0}\right),
\label{eq7}
\end{eqnarray}
with $\sigma_\mathrm{tot}=4\pi B_0$. Alternatively, this $f$ can be calculated with the coefficients of cosine polynomials $A_i$ (up to 4$^\mathrm{th}$
order) by
\begin{eqnarray}
f = 1 + \frac{A_2}{3A_0}+\frac{A_4}{5A_0},
\label{eq8}
\end{eqnarray}
with $\sigma_\mathrm{tot}=4\pi A_0 \times f$.
Here, Eqs.~5--8 are valid either in $c.m.$ or $lab.$ frame, and obviously $f$ is independent of the coordinate frame. For the $^{19}$F($p$,$\alpha_0$)$^{16}$O
reaction, the difference between $c.m.$ and $lab.$ differential cross sections is quite small, about 1\% at $\sim$90$^\circ$ in the energy region studied.
This difference can be neglected if compared to the uncertainty of experimental data.

One or two kinds of expansion coefficients were given in the previous works, and their relation was deduced in Ref.~\cite{iso56}. By using these coefficients,
we have plotted the factor $f$ in Fig.~\ref{fig9}. It shows that the factor $f$ assumes large values in correspondence of resonances, while non-resonant
region has a factor around unity. As a conclusion, to give an approximate estimate of the non-resonant part of the cross section, one could measure the
differential cross section at $\theta_\mathrm{lab}$=90$^\circ$, and then by multiplying a factor of 4$\pi$, the total cross section can be determined
(see Eq.~\ref{eq6}). This method can simplify the lengthy angular distribution measurements if one needs to know the behavior of the total cross section
far from a resonant peak. It is worthy of noting that there are still some discrepancies between different datasets as seen in Fig.~\ref{fig9}, which are
needed to be solved where necessary. In addition, it should be noted that the angular distribution factors ($f$) below 0.6 MeV are not ideally unity
(about 0.8$\sim$1.2), implying there are some resonances in this region which were actually observed by LOM15. This also demonstrates that the previous
non-resonance extrapolation set only the rough lower limits. Since there is a resonance around 0.113 MeV as shown in Fig.~\ref{fig8}, a future experiment
should measure either angular distribution or total cross section.

\begin{center}
\includegraphics[width=8.6cm]{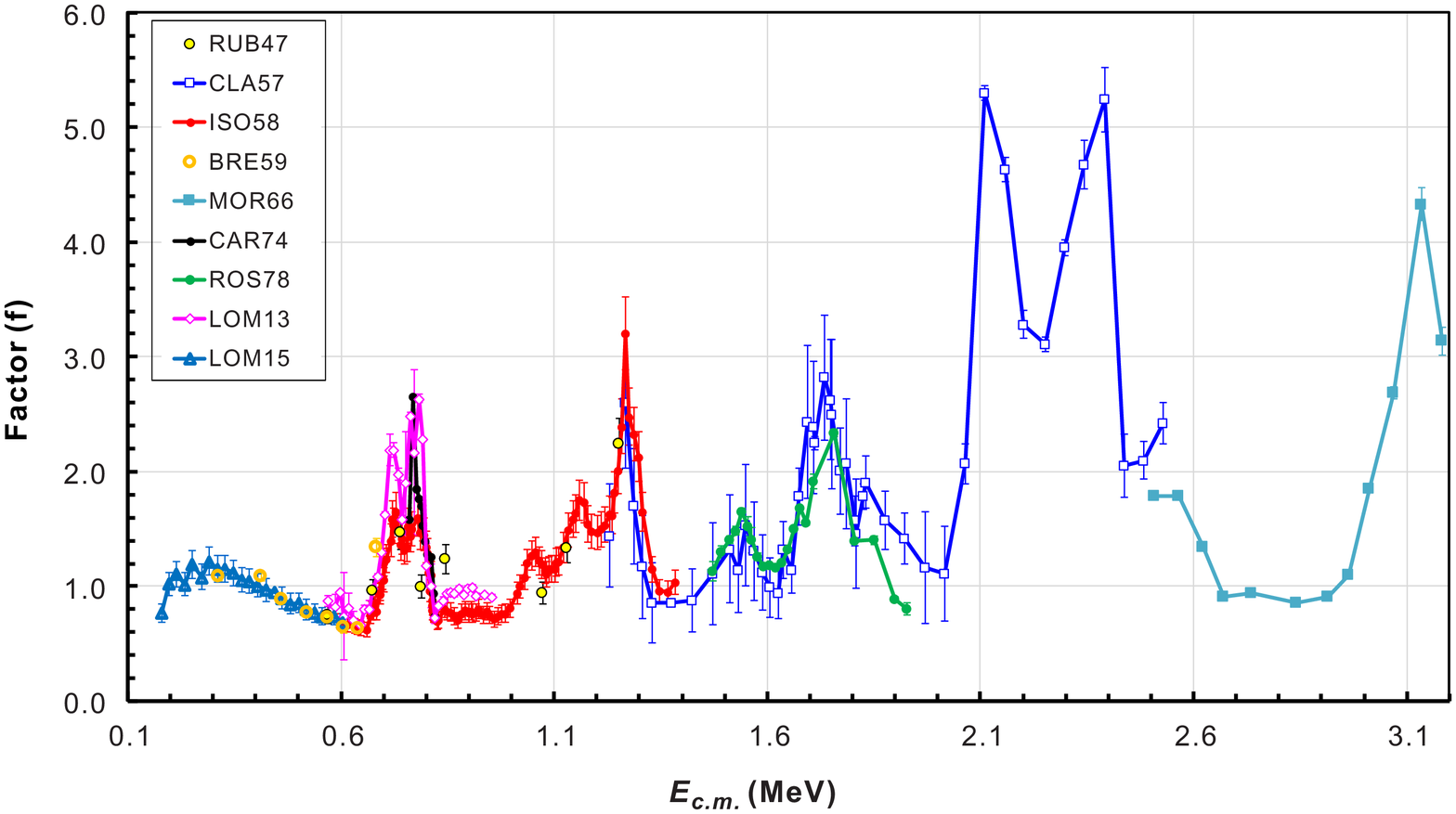}
\vspace{-2mm}
\figcaption{Angular distribution factor $f$ as a function of energy for the $^{19}$F($p$,$\alpha_0$)$^{16}$O reaction.}
\label{fig9}
\end{center}

\subsection{Differential cross section}
We have re-evaluated the differential cross sections observed at $\theta_\mathrm{lab}$=90$^\circ$ as shown in Fig.~\ref{fig10}. Here, the $E_{c.m.}$ energy
scale has been corrected for the energy loss in the target. For the differential cross section $d\sigma/d\Omega(90^\circ$), DIE80~\cite{die80} obtained an
absolute measured value of (1.05$\pm$0.09) mb/sr at $E_p$=1.354 MeV (with $\sim$1 keV target energy loss), while LER69~\cite{ler69}, in a dedicated series of
experiments, obtained absolute measured value of (1.02$\pm$0.10) mb/sr at $E_p$=1.360 MeV (with $\sim$7--24 keV target energy loss). Actually these peaks are
due to the same resonance after taking the target energy-loss effect into account, and they give rise to the peak at $E_{c.m.}$=1.280 MeV shown in Fig.~\ref{fig10}.
Because that there are no other available absolute measurements in this energy region, we adopted here the DIE80 excitation function as the reference.
The ISO58 and RAN58~\cite{ran58} data have been normalized to DIE80 with factors of 0.5 and 0.7, respectively. It shows that ISO58 and DIE80 data are
consistent down to about 0.8 MeV, below which they behave quite differently.

\begin{center}
\includegraphics[width=8cm]{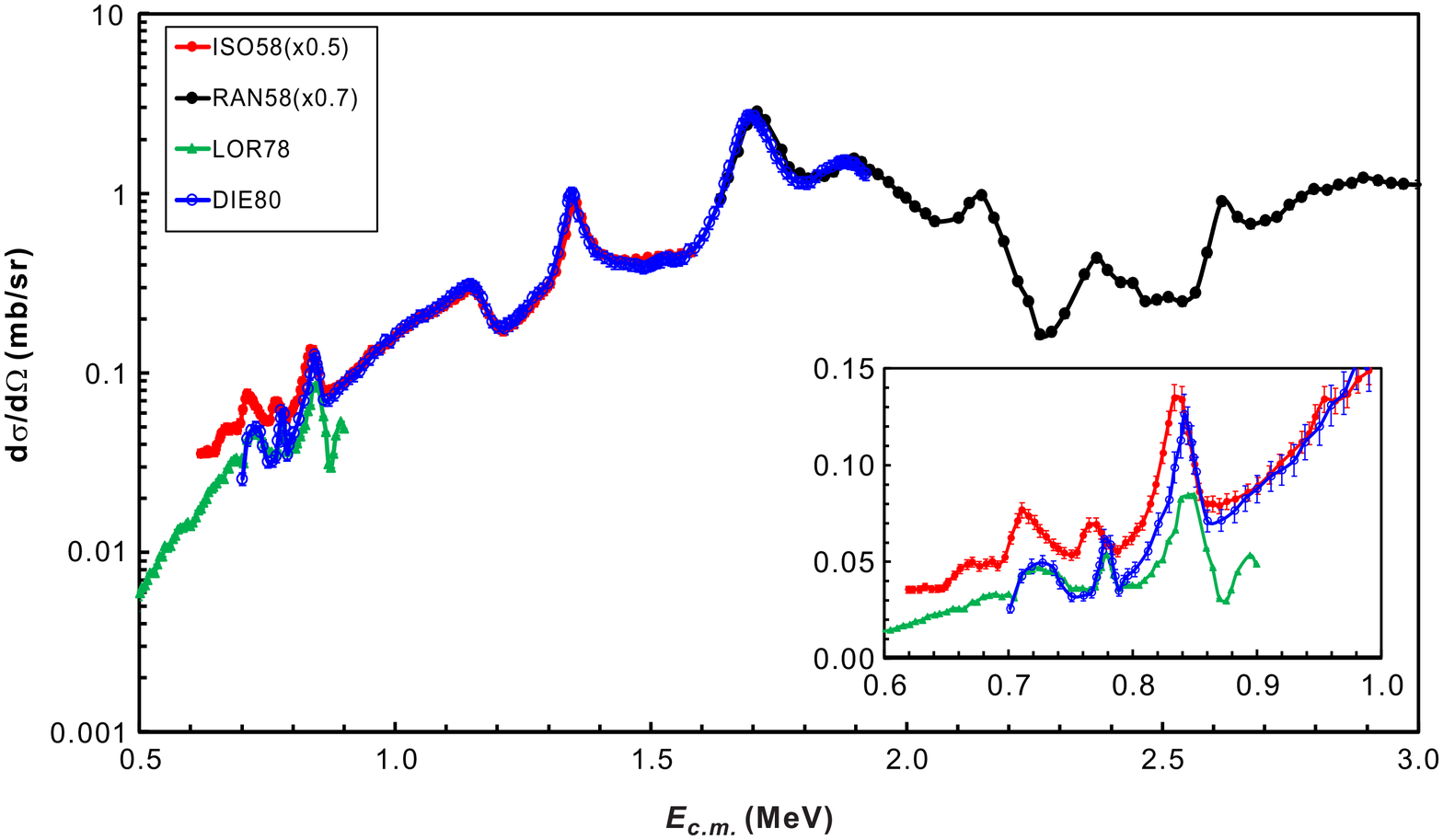}
\vspace{2mm}
\figcaption{Evaluation of $^{19}$F($p$,$\alpha_0$)$^{16}$O differential cross sections observed at $\theta_\mathrm{lab}$=90$^\circ$. Here,
``ISO58($\times$0.5)" are obtained by multiplying a correction factor of 0.5 on the digitized Fig.~2 data in Ref.~\cite{iso58}, ``RAN58($\times$0.7)"
by a factor of 0.7 on the digitized Fig. 2 data in Ref.~\cite{ran58}, and ``DIE80" as the reference discussed in the text. The original unpublished LOR78
data are shown for comparison. The enlarged small figure is inserted for clarity (in linear scale).}
\label{fig10}
\end{center}

As mentioned above, a normalization factor of 0.5 is adopted for the observed ISO58 data (i.e., Fig.~2 data in Ref.~\cite{iso58}). The rationality of this
normalization factor will be explained below. Firstly, we extracted the coefficients ($b_0$, $b_2$ and $b_4$) of the Legendre polynomials from Fig.~4
in ISO58, and then calculated the angular distribution factor $f$ by using Eq.~\ref{eq7}, and finally we calculated the differential cross sections by the
following relation as discussed above:
\begin{eqnarray}
\frac{d\sigma}{d\Omega}(90^\circ) =  \frac{1}{4\pi f} \times \left (\frac{2.97\times 10^{-4}\times b_0}{E_{c.m.}} \times 0.8 \right),
\label{eq9}
\end{eqnarray}
where the term in the parenthesis represents the total cross section with a normalization factor of 0.8 utilized in Sec.~3.1.1 for the ISO58 $S$-factor data.
Figure~\ref{fig11} shows the comparison between the two datasets. It shows that they are very consistent, and the normalized ISO58 data are consistent
very well with the DIE80 data (except the region below 0.8 MeV) as shown in Fig.~\ref{fig10}. In order to make both ISO58 $S$-factor and differential cross section
data consistent with other datasets simultaneously, the differential cross sections shown in Fig.~2 of ISO58 should be reduced by a factor of 0.5 (possibly due
to a mistake). In fact, this normalization factor of 0.5 is a kind of ``correction" factor.

In addition, the RAN58 derived a total $^{19}$F($p$,$\alpha_0$)$^{16}$O cross section value of 40 mb at the $E_p$=1.35 MeV resonance, based on the CLA57 angular
distribution. By considering the above normalization factor of 0.7, a value of 28 mb (=40$\times$0.7) is obtained, which is consistent with the present evaluated
value of (26.0$\pm$2.6) mb.

There are large discrepancies among ISO58, LOR78 and DIE80 data in the region below 0.8 MeV as shown in the inserted plot of Fig.~\ref{fig10}.
Roughly speaking, the LOR78 data are about a factor of 2 smaller than ISO58($\times$0.5), and we do not know the exact origin of such discrepancy. Here, the
unpublished LOR78 data haven't been included into the present evaluation. This underestimation is possibly owing to the target degradation, since LOM78 used
a very strong proton beam up to 200~$\mu$A. Recently, we have tested many CaF$_2$ and LiF targets, and found that the target degradation was very serious under
proton beam of about several $\mu$A~\cite{zha17}.
In addition, the exact reason why DIE80 is different from ISO58($\times$0.5) below about 0.84 MeV is also unknown. Here we assumed that it is again attributed to the
target degradation. In DIE80, it described that ``Beam currents were around 1~$\mu$A, on a 1~mm$^2$ spot”. At low energies the cross section becomes small, and
the machine time on the target should be longer than the higher energy region. This very sharp beam bombarding a very thin LiF target (5.3~$\mu$g/cm$^2$ of F)
during the long run could degrade the target seriously, and that's possibly why the DIE80 differential cross section reduced considerably. In contrast to the
ISO58 experiment, the proton beam bombarded a 50~$\mu$g/cm$^2$ CaF$_2$ target with currents of 0.4 to 2 $\mu$A, where a beam defining slit of 3.3 mm was utilized
``to insure durability of the target under the ion bombardment by reducing the current density". The much thicker CaF$_2$ target and reduced current density could
alleviate the impact of target degradation on the results.

\begin{center}
\includegraphics[width=8cm]{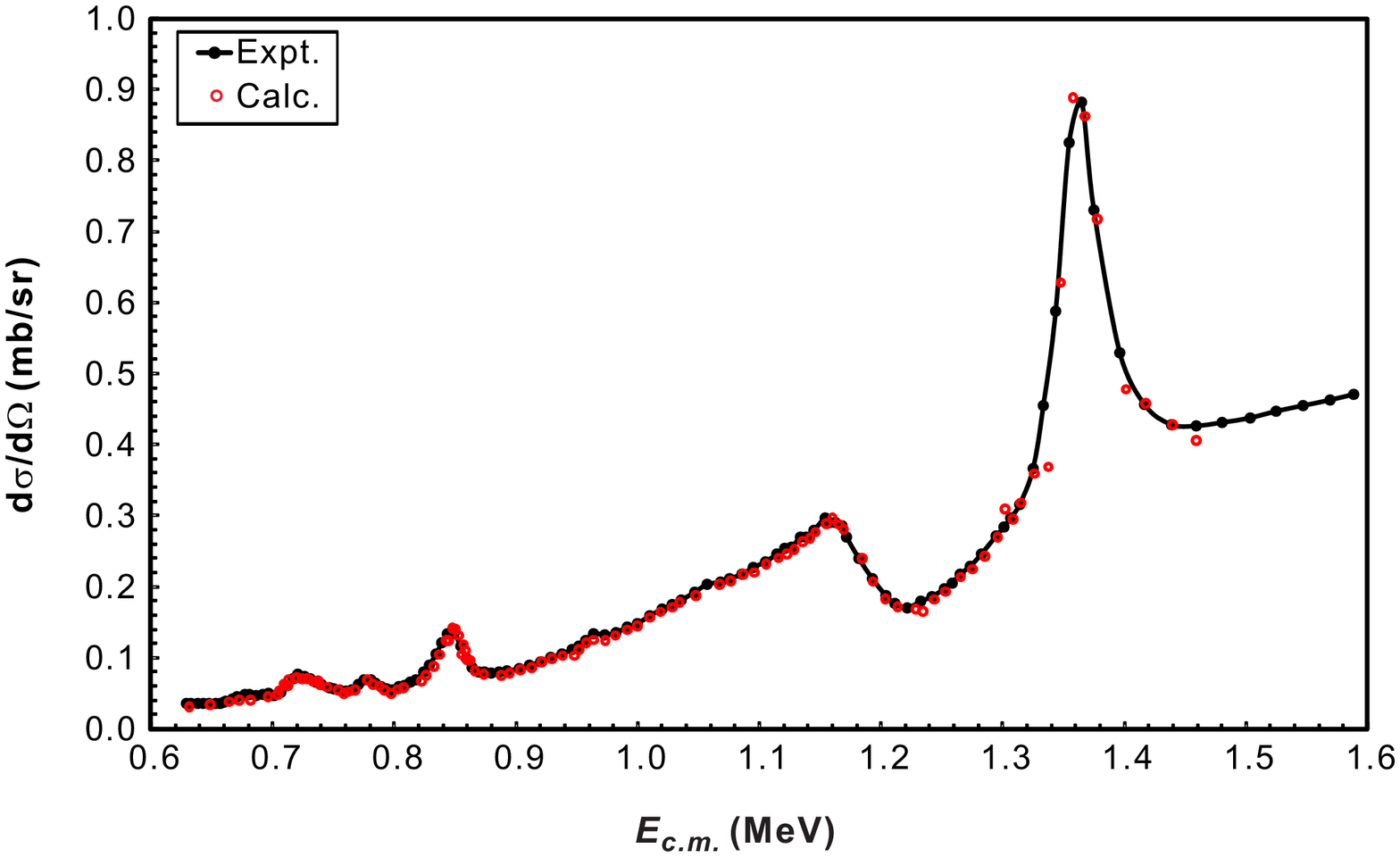}
\vspace{2mm}
\figcaption{Differential cross sections of the $^{19}$F($p$,$\alpha_0$)$^{16}$O reaction at $\theta_\mathrm{lab}$=90$^\circ$ evaluated based on the ISO58
data~\cite{iso58}. Here, ``Expt." represents exactly that of ``ISO58($\times$0.5)" shown in Fig.~\ref{fig10}, and ``Calc." represents the calculated one by
Eq.~\ref{eq9} as explained in the text.}
\label{fig11}
\end{center}

\section{Reaction rates}
It is well-known that the reaction rate of charged-particle induced reaction can be calculated, in terms of astrophysical $S$ factor, by the following
equation~\cite{rol88,ang99},
\end{multicols}
\begin{eqnarray}
N_A\langle \sigma v \rangle = N_A\left (\frac{8}{\pi \mu}\right)^{1/2}\frac{1}{(kT)^{3/2}}\int^{\infty}_{0}S(E) \mathrm{exp} \left[-\frac{E}{kT}-2\pi\eta \right]dE.
\label{eq10}
\end{eqnarray}
\begin{multicols}{2}
\noindent
As already discussed in Eq.~\ref{eq4}, the reduced mass $\mu$ is in units of amu, and it enters into the exponential term in the above equation. In the
present work, $\mu$ is precisely calculated with proton mass of 1.007825u, and $^{19}$F mass of 18.998403u~\cite{wan17}. If one simply approximates proton
and $^{19}$F mass as 1u, and 19u, respectively, the calculated penetration factor of exp(-2$\pi\eta$) will be different from the precise one. Such an
impact is shown clearly in Fig.~\ref{fig12}, where the approximated factor is enhanced considerably in the low energy region. In other words, the approximation
of mass values can considerably affect the reaction rate in the low temperature region.

The thermonuclear $^{19}$F($p$,$\alpha_0$)$^{16}$O rate has been calculated by numerical integration of our evaluated $S$ factors with Eq.~\ref{eq10}.
We divided the evaluated $^{19}$F($p$,$\alpha_0$)$^{16}$O $S$-factor datasets into following three regions:
(1) in the low energy region where no experimental data are available, we adopt the theoretical $R$-matrix results of LOM15 as shown in Fig.~\ref{fig8}
(with assumed uncertainty of 20\%~\cite{lom15});
(2) in the higher energy region of $E_{c.m.}$=4--10 MeV, the NACRE `WA63b' data~\cite{war63} shown in Fig.~\ref{fig2} are adopted (with assumed uncertainty
of 20\%~\cite{ang99});
(3) in the energy region of $E_{c.m.}$=0.2--3.2 MeV, we adopt the evaluated data and associated errors in Fig.~\ref{fig4} except the BRE59 data (because
of their large uncertainties). It should be noted that there are discrepancies between CLA57 and CUZ80 data as shown in Fig.~\ref{fig4}, and hence we adopt
the average of the two datasets in the reaction rate calculations, although the maximum difference resulting in the rate is less than 9\% (smaller than
3\% below 2 GK). Additionally, we assumed a $\pm$2 keV uncertainty of the experimental $E_{c.m.}$ energies (shown in Fig.~\ref{fig4}) in the numerical
integration, but this uncertainty results in no more than 3\% uncertainty to the lower and upper limits. The numerical values of the present reaction rate
and the associated lower and upper limits are listed in Table~\ref{tab1_rate}. Finally the present rate is parameterized by the standard format of~\cite{rau00},
\end{multicols}
\begin{eqnarray}
N_A\langle\sigma v\rangle &=& \mathrm{exp}(51.8361-\frac{9.79933}{T_9}+\frac{315.811}{T_9^{1/3}}-366.895T_9^{1/3}+16.2212T_9-0.863T_9^{5/3}+210.485\ln{T_9}) \nonumber \\
                          &+& \mathrm{exp}(48.7403-\frac{0.031187}{T_9}-\frac{11.441}{T_9^{1/3}}-32.2709T_9^{1/3}+3.34216T_9-0.2476T_9^{5/3}+8.72415\ln{T_9}) \nonumber \\
                          &+& \mathrm{exp}(6165.89-\frac{2.56546}{T_9}+\frac{759.439}{T_9^{1/3}}-9936.72T_9^{1/3}+6431.65T_9-5224.7T_9^{5/3}+1610.12\ln{T_9}) \, ,
\label{eq11}
\end{eqnarray}
\begin{multicols}{2}
\noindent
with a fitting error of less than 1.5\% over the entire temperature region of 0.007--10 GK.

\begin{center}
\includegraphics[width=7cm]{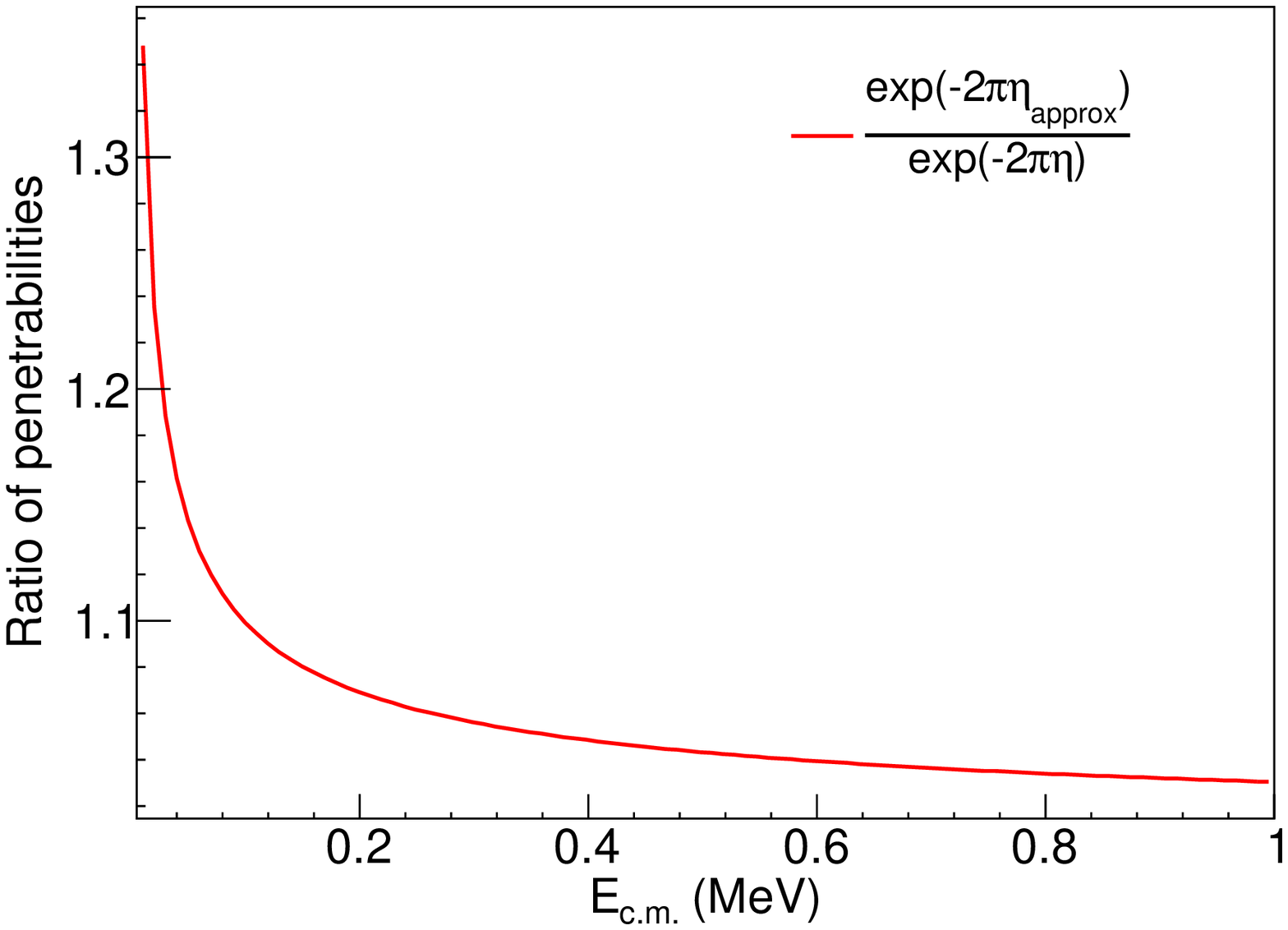}
\vspace{2mm}
\figcaption{Ratio of penetration factor exp(-2$\pi\eta$) by using precise and approximated Sommerfeld parameter $\eta$ for the $^{19}$F+$p$ system. Here
the parameters $\eta$ and $\eta_\mathrm{approx}$ are calculated respectively by using the precise and the approximate value of the reduced mass $\mu$.
See text for details.}
\label{fig12}
\end{center}

\begin{center}
\tabcaption{\label{tab1_rate} Thermonuclear $^{19}$F($p$,$\alpha_0$)$^{16}$O reaction rate associated with the lower and upper limits (in units of cm$^3$s$^{-1}$mol$^{-1}$).}
\footnotesize
\begin{tabular*}{85mm}{c@{\extracolsep{\fill}}cccc}
\toprule $T_9$  & Rate  & Lower limit  & Upper limit \\
\hline
0.007 &	6.161E-29 &	4.929E-29 &	7.393E-29 \\
0.008 &	3.490E-27 &	2.792E-27 &	4.189E-27 \\
0.009 &	1.058E-25 &	8.468E-26 &	1.270E-25 \\
0.010 &	1.999E-24 &	1.599E-24 &	2.399E-24 \\
0.011 &	2.610E-23 &	2.088E-23 &	3.132E-23 \\
0.013 &	1.937E-21 &	1.549E-21 &	2.324E-21 \\
0.015 &	6.409E-20 &	5.127E-20 &	7.691E-20 \\
0.018 &	4.353E-18 &	3.482E-18 &	5.224E-18 \\
0.020 &	4.434E-17 &	3.548E-17 &	5.321E-17 \\
0.025 &	4.645E-15 &	3.716E-15 &	5.574E-15 \\
0.030 &	1.618E-13 &	1.295E-13 &	1.942E-13 \\
0.040 &	2.897E-11 &	2.318E-11 &	3.477E-11 \\
0.050 &	1.185E-09 &	9.481E-10 &	1.422E-09 \\
0.060 &	2.034E-08 &	1.627E-08 &	2.441E-08 \\
0.070 &	1.948E-07 &	1.558E-07 &	2.338E-07 \\
0.080 &	1.224E-06 &	9.790E-07 &	1.470E-06 \\
0.090 &	5.639E-06 &	4.505E-06 &	6.773E-06 \\
0.100 &	2.064E-05 &	1.647E-05 &	2.481E-05 \\
0.110 &	6.371E-05 &	5.076E-05 &	7.666E-05 \\
0.140 &	9.585E-04 &	7.643E-04 &	1.153E-03 \\
0.180 &	1.368E-02 &	1.112E-02 &	1.623E-02 \\
0.200 &	3.932E-02 &	3.241E-02 &	4.622E-02 \\
0.250 &	3.219E-01 &	2.738E-01 &	3.699E-01 \\
0.300 &	1.560E+00 &	1.358E+00 &	1.763E+00 \\
0.350 &	5.386E+00 &	4.762E+00 &	6.010E+00 \\
0.400 &	1.470E+01 &	1.314E+01 &	1.627E+01 \\
0.450 &	3.390E+01 &	3.049E+01 &	3.730E+01 \\
0.500 &	6.891E+01 &	6.225E+01 &	7.557E+01 \\
0.600 &	2.184E+02 &	1.980E+02 &	2.388E+02 \\
0.700 &	5.446E+02 &	4.935E+02 &	5.958E+02 \\
0.800 &	1.159E+03 &	1.048E+03 &	1.270E+03 \\
0.900 &	2.200E+03 &	1.984E+03 &	2.415E+03 \\
1.000 &	3.833E+03 &	3.450E+03 &	4.216E+03 \\
1.250 &	1.183E+04 &	1.060E+04 &	1.304E+04 \\
1.500 &	2.867E+04 &	2.560E+04 &	3.166E+04 \\
1.750 &	5.973E+04 &	5.303E+04 &	6.602E+04 \\
2.000 &	1.115E+05 &	9.831E+04 &	1.233E+05 \\
2.500 &	3.011E+05 &	2.623E+05 &	3.330E+05 \\
3.000 &	6.287E+05 &	5.434E+05 &	6.947E+05 \\
3.500 &	1.094E+06 &	9.407E+05 &	1.207E+06 \\
4.000 &	1.672E+06 &	1.434E+06 &	1.842E+06 \\
5.000 &	3.037E+06 &	2.599E+06 &	3.338E+06 \\
6.00  &	4.506E+06 &	3.855E+06 &	4.948E+06 \\
7.00  &	5.898E+06 &	5.045E+06 &	6.475E+06 \\
8.00  &	7.156E+06 &	6.120E+06 &	7.860E+06 \\
9.00  &	8.257E+06 &	7.060E+06 &	9.079E+06 \\
10.00 &	9.203E+06 &	7.865E+06 &	1.013E+07 \\
\bottomrule
\end{tabular*}
\vspace{0mm}
\end{center}
\vspace{0mm}

The comparison between different rates relative to the present rate is shown in Fig.~\ref{fig13}. The difference among LOM15, IND17 and NACRE reaction
rates was already discussed before, and will not be repeated here. Fig.~\ref{fig13} shows that our rate is lower than all the previous rates above
$\sim$1 GK, owing to the present smaller evaluated ISO58 and CLA57 $S$ factors. Within the large uncertainties ((10$\sim$20)\% for the present,
20\%~\cite{lom15} for LOM15 and 16\%~\cite{ind17} for IND17), our rate is consistent with the LOM15 and IND17 rates, but it is larger than the NACRE one
when below 1 GK (where a small non-resonant $S$-factor was assumed in the low energy region). Furthermore, Fig.~\ref{fig13} shows that our rate is
smaller than the IND17 rate in the low temperature region (e.g., by up to about 20\% around 0.007 GK). Since the low energy part of the $S$-factor quoted
in IND17 is quite similar to the present one, we believe that the main source leading to the disagreement between the present and IND17 data sets at very
low $T_9$ values could be a rough approximation of the reduced mass value in IND17. In this context, it is worth noting that effects due to use of an
approximated reduced mass value are almost canceled out when one reports ratio of reactions rates calculated under the same approximation. In the
temperature region of 0.007$\sim$1 GK, our rate is almost identical to that of LOM15 since we adopted the similar $S$ factors at low energies. The small
differences originate from the fact that we adopt the experimental $S$-factor data at energies below 0.8 MeV, while LOM15 adopted the $R$-matrix predictions
in the same energy region. This is why one may see a small bump (about 8\%) around 0.2 GK in Fig.~\ref{fig13} due to a bump structure observed around
0.185 MeV shown in Fig.~\ref{fig8}, where no such structure was predicted by the LOM15's $R$-matrix calculation. The uncertainties of the present low
temperature rate are estimated to be $\sim$20\%, which are mainly determined by the large uncertainties adopted for the $R$-matrix calculations
(20\% assumed in Ref.~\cite{lom15}) and those of the experimental data.

\begin{center}
\includegraphics[width=8.6cm]{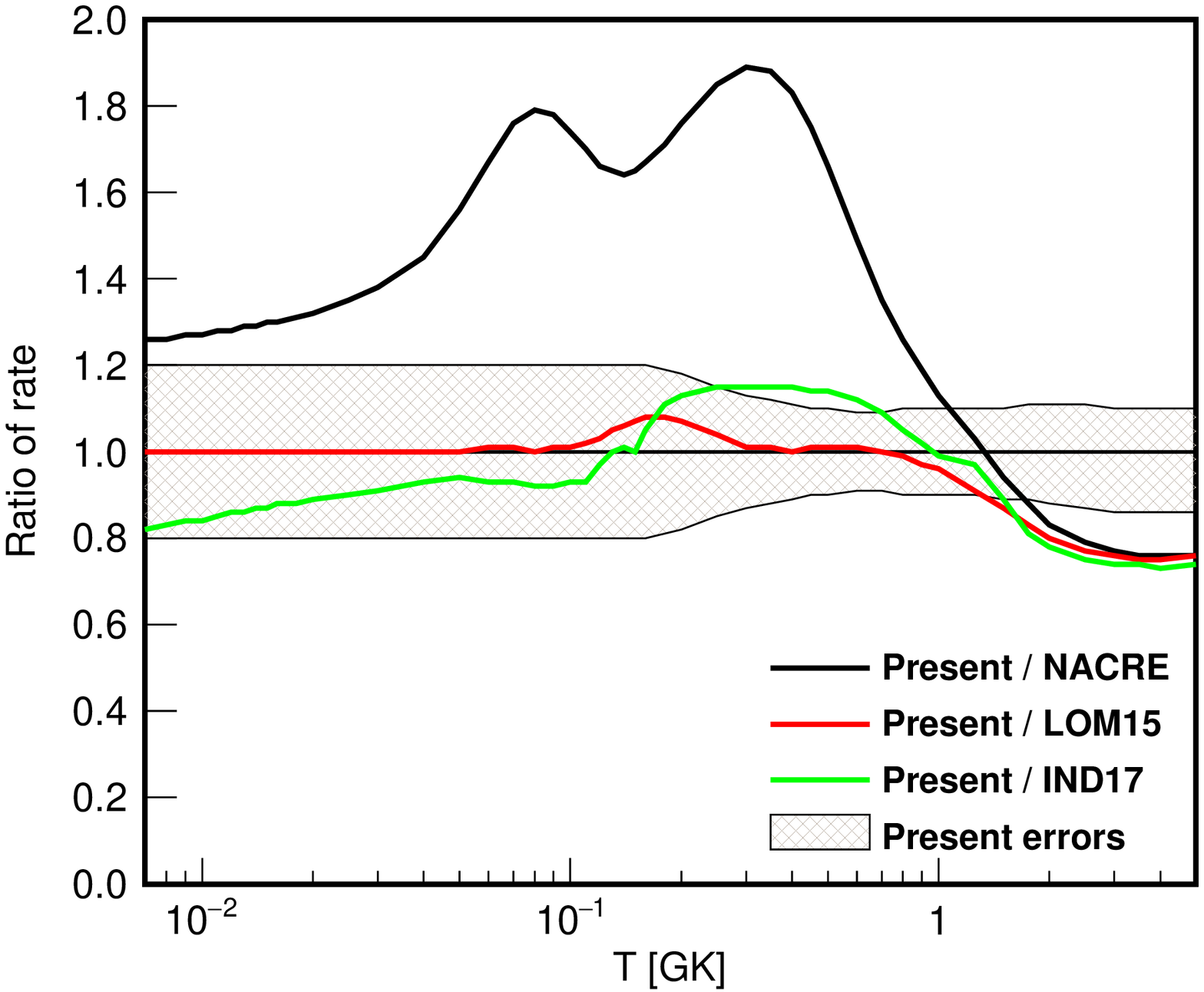}
\vspace{2mm}
\figcaption{$^{19}$F($p$,$\alpha_0$)$^{16}$O reaction rate ratios between the present and NACRE~\cite{ang99}, LOM15~\cite{lom15}, IND17~\cite{ind17} rates.
The associated error of the present one is shown as the gridded band.}
\label{fig13}
\end{center}

\section{Summary and outlook}
We have re-evaluated the available astrophysical $S$ factors of $^{19}$F($p$,$\alpha_0$)$^{16}$O reaction in the energy region of $E_{c.m.}$=0.2--3.2 MeV.
A thermonuclear $^{19}$F($p$,$\alpha_0$)$^{16}$O reaction rate in the temperature region of 0.007--10 GK has been calculated based on these evaluated data
and the low-energy theoretical $R$-matrix extrapolation. It shows that our new rate is smaller than the previous one~\cite{ind17} at temperatures below
$\sim$0.2 GK, e.g., by up to about 20\% around 0.01 GK; this effect seems to be due to an approximation utilized in the previous numerical integration.
Furthermore, our rate is smaller at temperatures above $\sim$1 GK, e.g., by about 20\% around 1.75 GK, mainly because we have re-evaluated the previous
data of Ref.~\cite{iso58}, which had not been interpreted correctly in the previous NACRE compilation. The present interpretation is supported by direct
experimental data. However, the ($p$,$\alpha_\gamma$) channel dominates the total rate in the temperature above $\sim$0.2 GK, and hence such lowering in
the ($p$,$\alpha_0$) rate does not change appreciably the total rate. The present rate uncertainties are still large, about 20\% in the low temperature
region of 0.007--0.2 GK, where the ($p$,$\alpha_0$) channel dominates the total $^{19}$F($p$,$\alpha$)$^{16}$O rate. This temperature region corresponds
to an energy $E_{c.m.}$ below $\sim$240 keV, where the precise experimental cross section (or $S$ factor) data are strongly required for astrophysical
nucleosynthesis studies in AGB stars. In addition, we find a considerably large discrepancy of the 90$^\circ$ differential cross sections between different
works below 0.9 MeV, which also needs further experimental clarification.

In 2014, the National Natural Science Foundation of China (NSFC) approved the Jinping Underground Nuclear Astrophysics laboratory (JUNA) project~\cite{liu16},
which aims at direct cross-section measurements of four key stellar nuclear reactions right down to the Gamow windows. In order to solve the observed
fluorine overabundances in AGB stars, measuring the key $^{19}$F($p$,$\alpha$)$^{16}$O reaction at effective burning energies (i.e., at Gamow window of
$E_{c.m.}$=70--350) has been established as one of the scientific research sub-projects~\cite{hjj16}, with the sufficient accuracy required by the stellar
model calculations. We hope that the new direct experimental data will help people to expound the element abundances problem as well as the heavy-element
nucleosynthesis scenario, by putting various astrophysical models on a firmer experimental ground.



\vspace{2mm}
\centerline{\rule{80mm}{0.1pt}}
\vspace{2mm}


\end{multicols}

\clearpage
\end{CJK*}
\end{document}